\documentclass[aps,pre,twocolumn,amssymb,amsmath,showpacs,showkeys,superscriptaddress,floatfix]{revtex4}

\usepackage[final]{graphicx}

\usepackage{dcolumn}

\begin{document}

\title{Properties of Quantum Systems via Diagonalization\\
of Transition Amplitudes I: Discretization Effects}
\author{Ivana Vidanovi\'c}
\author{Aleksandar Bogojevi\'c}\email[E-mail: ]{aleksandar.bogojevic@scl.rs}
\author{Aleksandar Beli\'c}
\affiliation{Scientific Computing Laboratory, Institute of Physics Belgrade, Pregrevica 118, 11080 Belgrade, Serbia}
\homepage[Home page: ]{http://www.scl.rs/}

\begin{abstract}
We analyze the method for calculation of properties of non-relativistic quantum systems based on exact diagonalization of space-discretized short-time evolution operators. In this paper we present a detailed analysis of the errors associated with space discretization. Approaches using direct diagonalization of real-space discretized Hamiltonians lead to polynomial errors in discretization spacing $\Delta$. Here we show that the method based on the diagonalization of the short-time evolution operators leads to substantially smaller discretization errors, vanishing exponentially with $1/\Delta^2$. As a result, the presented calculation scheme is particularly well suited for numerical studies of few-body quantum systems. The analytically derived discretization errors estimates are numerically shown to hold for several models. In the followup paper \cite{pqseeo2} we present and analyze substantial improvements that result from the merger of this approach with the recently introduced effective-action scheme for high-precision calculation of short-time propagation.
\end{abstract}

\preprint{SCL preprint}
\maketitle

\section{Introduction}
\label{sec:intro}

In the standard operator formulation of quantum mechanics, the description of a physical system is based on constructing the Hamiltonian operator $\hat{H}$. Properties of quantum systems are obtained by solving the corresponding Schr\" odinger equation,
\begin{equation}
\hat{H}|\psi\rangle=E|\psi\rangle\, .
\label{eq:sch}
\end{equation}
Exact solutions can be found only for a very limited set of simple models. A wide variety of analytical approximation techniques has been developed in the past for treatment of such problems. In addition, the last two decades have seen a rapid growth in the application of different numerical methods for solving the Schr\" odinger equation. In this paper we focus on approaches based on real-space discretization, which usually start from some given finite-difference prescription. Such methods have been extensively studied in the past, and the main difficulties follow from the finite-difference representations of the kinetic operator.

A numerical approach based on diagonalizing of the evolution operator, introduced in Ref.~\cite{sethia}, does not suffer from problems with the representation of differential operators on real-space grids, and has substantial advantages in practical applications to few-body problems. Effectively, in this way the problem is transfered from that of representing the kinetic operator on a real-space grid to the calculating of corresponding transition amplitudes. Detailed analysis of the errors associated with the implementation of this approach has not been presented before, and is the main result of the current paper. It provides full understanding of the method and allows its optimal use, as well as further significant improvements within a generalized calculation scheme, presented in our followup paper \cite{pqseeo2}.

The advantages of the method discussed in this paper \cite{sethia, sethiacpl1, sethiajcp, sethiacpl2} follow from two key properties. First, the objects being diagonalized are transition amplitudes, which are well defined irrespective of discretization scheme, i.e. the exponential of the Hamiltonian effectively regularizes the kinetic operator, making possible representations of the evolution operator that do not depend on the space grid. Second, the successful diagonalization of the evolution operator $\exp(-t\hat{H})$ for any time of propagation $t$ immediately gives the solution of the eigenproblem for the Hamiltonian. Thus, the time of propagation in this approach is just an auxiliary parameter. Said another way, we use the time-dependent evolution operator to extract time-independent information regarding the quantum system. If one could calculate transition amplitudes exactly, then the obtained results for the energy eigenproblem would not depend on the time of propagation. However, in practical applications one uses some approximation scheme to calculate the amplitudes, and in this case the precision of the obtained results for energy eigenvalues and eigenstates does depend on time $t$. The general applicability of the outlined method follows from the fact that one can use short-time propagation amplitudes to obtain high accuracy results.

In order to complete this numerical method and make it generally applicable, it is necessary to address the following key questions:
\begin{enumerate}
\item
How to analytically estimate the effects of spatial discretization?
\item
How to optimize the choice of evolution time $t$, so as to minimize errors?
\item
How to accurately calculate transition amplitudes?
\end{enumerate}
The authors in Ref.~\cite{sethia} have only briefly commented on the first two questions, and numerically determined the values of parameters that can be used for precise calculations of energy eigenvalues and eigenstates for several models. To numerically calculate transition amplitudes, they exclusively relied on the naive short-time approximation formula,
\begin{equation}
 \langle x|e^{-t \hat{H}}|y\rangle\approx\frac{1}{\sqrt{2\pi t}}\, e^{-\frac{(x-y)^2}{2 t}-t \frac{V(x)+V(y)}{2}}\, ,
\label{eq:naive}
\end{equation}
which is correct only to order $O(t)$ for energy eigenvalues.

In this paper we address the above questions, which have not been fully answered before. In Section \ref{review} we present the method and notation, and identify the sources of the errors present in real-space discretization approaches. In Section \ref{body} we analyze in detail questions 1 and 2, and discuss the effects of discretization on the obtained properties of physical systems. We analytically derive estimates for errors coming from space discretization coarseness, finite size effects, and choice of evolution time parameter $t$. All the analytically derived results are numerically verified to hold on several instructive models.

The followup paper \cite{pqseeo2} continues this investigation and significantly improves the method by applying the recently introduced effective action approach \cite{b, prl-speedup, prb-speedup, pla-euler, pla-manybody} to completely resolve the problem formulated in question 3. This has been addressed recently \cite{k1, k2, k3} using various approaches. We stress that use of higher-order effective actions represents an efficient and numerically inexpensive way to calculate transition amplitudes that lead to many orders of magnitude increase in precision of calculated properties of the system.

The expressions written throughout this paper are, for compactness of notation, for one particle in one dimension. Extension to more particles and dimensions is straightforward, just as with the above short-time transition amplitude. Note that we are working in imaginary time, which is well suited for numerical calculations and does not affect in any way energy levels nor other time-independent properties of the system. We have also set $\hbar$ to unity.

\section{Space-discretized Schr\"odinger equation}
\label{review}

In the coordinate representation the time-independent Schr\" odinger's equation takes the form
\begin{equation}
 \int \mathrm{d}y\,\langle x|\hat{H}|y\rangle\, \langle y|\psi\rangle = E\, \langle x|\psi\rangle\, .
\end{equation}
The standard way to numerically implement exact diagonalization is to go from continuous coordinates $x$ to ones living on a discrete space grid $x_n=n\Delta$, where $\Delta$ is a given spacing and $n\in \mathbb{Z}$. Integrations in the above equation are performed using the simple rectangular quadrature rule, or some higher-order finite-difference formula. This completes the transition to the space-discretized counterpart of the continuous theory, however, to represent this on a computer we still have to restrict the integers $n$ to a finite range. This is equivalent to introducing a space cutoff $L$, or putting the system in a infinitely high potential box. For example, the rectangular quadrature rule leads to the following space-discretized Schr\" odinger equation
\begin{equation}
 \sum_{m=-N}^{N-1}H_{nm}\langle m\Delta|\psi\rangle=E(\Delta, L)\, \langle n \Delta|\psi\rangle\, ,
\label{eq:hamdis}
\end{equation}
where $H_{nm}=\Delta\cdot \langle n\Delta|\hat{H}|m\Delta \rangle$, $N=[L/\Delta]$, and square brackets represent the integer part of the argument. As a result, we have obtained a $2N\times 2N$ matrix that represents the Hamiltonian of the system. The eigenvalues of this matrix depend on the two parameters introduced in the above discretization process: cutoff $L$ and discretization step $\Delta$. Continuous physical quantities are recovered in the limit $L\to \infty$ and $\Delta\to 0$. The outlined procedure is very useful in dealing with spatially localized physical problems, such as electronic structure calculations in semiconductor and polymer physics \cite{spacedisrev}.

The two approximations involved in the discretization procedure, characterized by parameters $\Delta$ and $L$, are common steps in solving eigenproblems of Hamiltonians, and as such have been extensively analyzed. The imposed constraint on the values of spatial coordinates to the finite interval $(-L,L)$ is a valid approach for capturing information on localized eigenstates. In this approximation the system is effectively surrounded  by an infinitely high wall, and as the cutoff $L$ tends to infinity, we approach the exact energy levels always from above \cite{ajp1, ajp2}, which is a typical variational behavior. Therefore, we designate errors associated with the cutoff $L$ as variational. The effects of the discretization step $\Delta$ are much more complex, and follow from the fact that the kinetic energy operator cannot be exactly represented on finite real-space grids. For example, a typical naive discretization of the kinetic energy operator gives in our notation the following Hamiltonian matrix elements \cite{d}
\begin{equation}
H_{nm}=\left\{
\begin{array}{ll}
 1/\Delta^2+V(n \Delta)& \mbox{if } n=m\\
-1/(2 \Delta^2)& \mbox{if } |n-m|=1\\
0\quad &\mbox{otherwise.}
\end{array}
\right.
\label{eq:tb}
\end{equation}
Note that in the absence of a potential term $V$ in the Hamiltonian, the above definition corresponds to a tight-binding model \cite{d}. This prescription leads to numerical results for eigenvalues which in the $\Delta\to 0$ limit converge to the exact continuum values as $\Delta^2$. The errors associated with this approach have non-variational behavior, i.e. the obtained results are not always upper bounds of the exact energy levels. Several papers discuss this issue and analyze the behavior of errors in the direct diagonalization approach (for more details, see Refs.~\cite{spacedis1, spacedis2, skylaris} and references therein). The state-of-the-art in this approach is a set of systematically improved prescriptions for discretization of the kinetic energy operator, which speeds up convergence to the continuum limit to higher powers of $\Delta^2$. However, within this approach convergence is always polynomial in $\Delta$. Some recent results \cite{spacedis2, skylaris} also exist on extensions of this approach that provide effective variational behavior of the discretized kinetic energy operator.

\begin{figure}[!t]
\centering
\includegraphics[width=8.5cm]{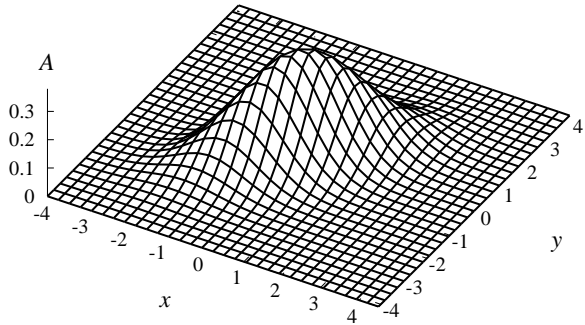}
\caption{Harmonic oscillator transition amplitude as a function of coordinates $x$ and $y$ for $t=1$, frequency $\omega=1$ and mass $M=1$.}
\label{fig:3dlhoamplitude}
\end{figure}
As outlined in the Introduction, in this paper we focus on an alternative approach, based on solving the eigenproblem of the corresponding transition amplitudes as proposed in \cite{sethia}. The central equation is
\begin{equation}
 \sum_{m=-N}^{N-1} A_{nm}(t)\, \langle m\Delta|\psi\rangle=e^{-t\, E(\Delta,L,t)}\, \langle n \Delta|\psi\rangle\, ,
\label{eq:evodis}
\end{equation}
where $A_{nm}(t)=\Delta\cdot A(n \Delta, m\Delta;t)=\Delta\cdot \langle n\Delta|e^{-t\hat{H}}|m\Delta\rangle$. In this approach the time of evolution $t$ plays the role of an auxiliary parameter. This parameter is not related to the discretization, and in a continuous theory it does not affect the obtained eigenvalues and eigenstates. However, in a discretized theory the numerically calculated eigenvalues and eigenstates will necessarily depend on this parameter as well, as emphasized by the right-hand size of Eq.~(\ref{eq:evodis}).
Therefore, the original problem is now transformed into the eigenproblem of the matrix $A_{nm}(t)$, whose indices take all integer values in the range
$-N\leq n,m< N$, where $N=[L/\Delta]$.

Fig.~\ref{fig:3dlhoamplitude} shows how a typical transition amplitude, in this case that of a harmonic oscillator, depends on coordinates $x$ and $y$. As can be seen from the figure, transition amplitudes are spatially well localized. This is particularly simple to understand for the short times of propagation that we consider. In this case the kinetic term exponentially localizes the transition amplitude matrix to the vicinity of the main diagonal. Similarly, the potential brings about exponential localization along the main diagonal around its minimum. The localization of dominant values of the transition amplitude to a small area in the $x-y$ plane gives practical justification for introduction of space cutoff $L$ in this approach.

\begin{figure}[!h]
\centering
\includegraphics[width=8.5cm]{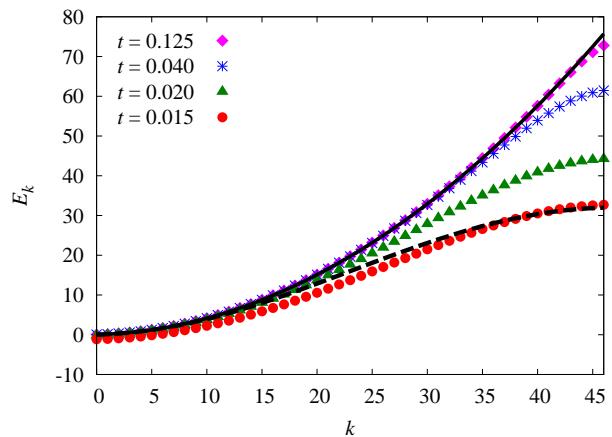}
\caption{(Color online) Eigenspectrum of a free particle in a box. Eigenvalues $E_k$ are given as a function of level number $k$. The solid line gives the exact parabolic dispersion $E_k=\pi^2(k+1)^2/8L^2$, while the dashed line presents results calculated in the tight-binding approximation. The graph also shows numerical results obtained by the diagonalization of transition amplitudes for different values of time of evolution $t$. All the numerical calculations are for $L=6$ and $\Delta=0.25$, hence $N=L/\Delta=24$.}
\label{fig:freeeigenspectrum}
\end{figure}

In continuum theory, the transition amplitude eigenproblem is mathematically equivalent to the Schr\" odinger equation. It is important to stress, however, that the procedure of space discretization introduces important differences between eigenproblems (\ref{eq:hamdis}) and (\ref{eq:evodis}).
In particular, as we will show in the next section, the procedure based on the diagonalization of transition amplitudes leads to much faster (non-polynomial) convergence. An illustration of the relation of these two calculation schemes is shown in Fig.~\ref{fig:freeeigenspectrum} which compares the exact parabolic dispersion of a free particle in a box with numerical calculations based on the diagonalizations of the Hamiltonian and of the transition amplitudes. From the figure we see that the time parameter $t$ in the transition amplitude approach plays an important role. Increase of $t$ gives better agreement with the exact dispersion relation.

\section{Diagonalization of space-discretized transition amplitudes}
\label{body}

The free particle transition amplitude
\begin{equation}
A^{free} (x,y;t)=\frac{1}{\sqrt{2\pi t}}\, \exp\left(-\frac{(x-y)^2}{2t}\right)
\end{equation}
satisfies
\begin{equation}
\label{int}
\int \mathrm{d}x\, A^{free} (x,y;t)=1\ .
\end{equation}
The consequence of this is conservation of probability. In the space-discretized analogue of this model $x=n\Delta$, $y=m\Delta$, and the transition amplitude is
$A^{free}_{nm}(t)=\Delta\,A^{free}(n\Delta,m\Delta;t)$. Using the Poisson summation formula
\begin{equation}
\label{poisson}
\sum_{n\in \mathbb{Z}}\exp\left(-\alpha n^2\right)=
\sqrt{\frac{\pi}{\alpha}}\,\sum_{n\in \mathbb{Z}}\exp\left(-\frac{\pi^2}{\alpha} n^2\right)\, ,
\end{equation}
we find that the space discretized free particle amplitude satisfies
\begin{equation}
\sum_{n\in \mathbb{Z}}A^{free}_{nm}(t)=\sum_{n\in \mathbb{Z}}e^{-\frac{2\pi^2}{\Delta^2} n^2 t}\approx
1+2\exp\left(-\frac{2\pi^2}{\Delta^2}t\right)\ .
\end{equation}
Conservation of probability is thus obtained only in the continuum limit $\Delta\to 0$. Note that the effect of discretization is non-perturbative in discretization step $\Delta$, i.e. it is smaller than any power of $\Delta$. The effect of discretization is also universal in that it holds for all models, since the free particle transition amplitude is the dominant term in the short time expansion of the transition amplitude of a general theory. 

To show this explicitly we use the short time expansion of the transition amplitude of a general theory \cite{b} to show that
\begin{equation}
\int \mathrm{d}x\, A (x,y;t)=\frac{1}{\sqrt{2\pi t}}\int \mathrm{d}x\,e^{-\frac{1}{2t}x^2}\sum_l t^lf_l(x,y)\ ,
\end{equation}
where $f_0=1$ and the other $f_l$ are given functions of the potential and its derivatives. Writing the even part of $f_l(x,y)$ as $g(x^2,y)$ we find
\begin{equation}
\int \mathrm{d}x\, A (x,y;t)=\frac{1}{\sqrt{t}}\sum_l t^lg_l(2t^2\partial_t,y)\sqrt{t}\ .
\end{equation}
Similarly, using the above Poisson summation formula, we find
\begin{eqnarray}
\lefteqn{
\sum_{n\in \mathbb{Z}}A_{nm}(t)-\int \mathrm{d}x\, A (x,y;t)=}\nonumber\\
&&=\frac{2}{\sqrt{t}}\sum_l t^lg_l(2t^2\partial_t,y)\sqrt{t}\exp\left(-\frac{2\pi^2}{\Delta^2}t\right)\ .
\end{eqnarray}
Performing the indicated differentiations the right hand side becomes $\exp(-\frac{2\pi^2}{\Delta^2}t)\cdot \sum_l h_l(y)t^l$. One could now calculate the $h_l$ from the short time expansions $f_l$. The $p$-level effective action gives a short time expansion that is truncated at order $t^p$. As a result $\sum_l h_l(y)t^l$ is a polynomial in time of order $p$. The dominant short time behavior is thus given by the universal exponential term. As a result the transition of a general model to its space discretized form is given by
\begin{equation}
\sum_{n\in \mathbb{Z}}A_{nm}(t)-\int \mathrm{d}x\, A (x,y;t)\sim\exp\left(-\frac{2\pi^2}{\Delta^2}t\right)\ .
\end{equation}

\begin{figure}[!b]
\centering
\includegraphics[width=8.5cm]{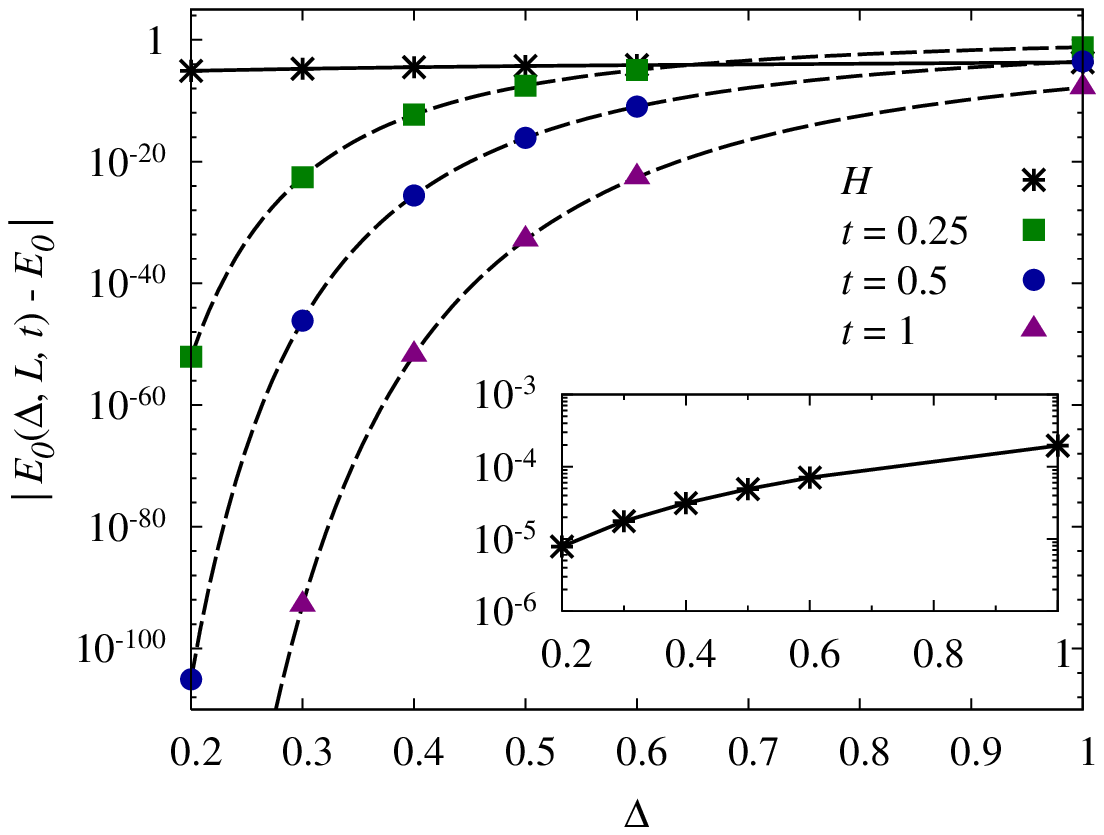}
\includegraphics[width=8.5cm]{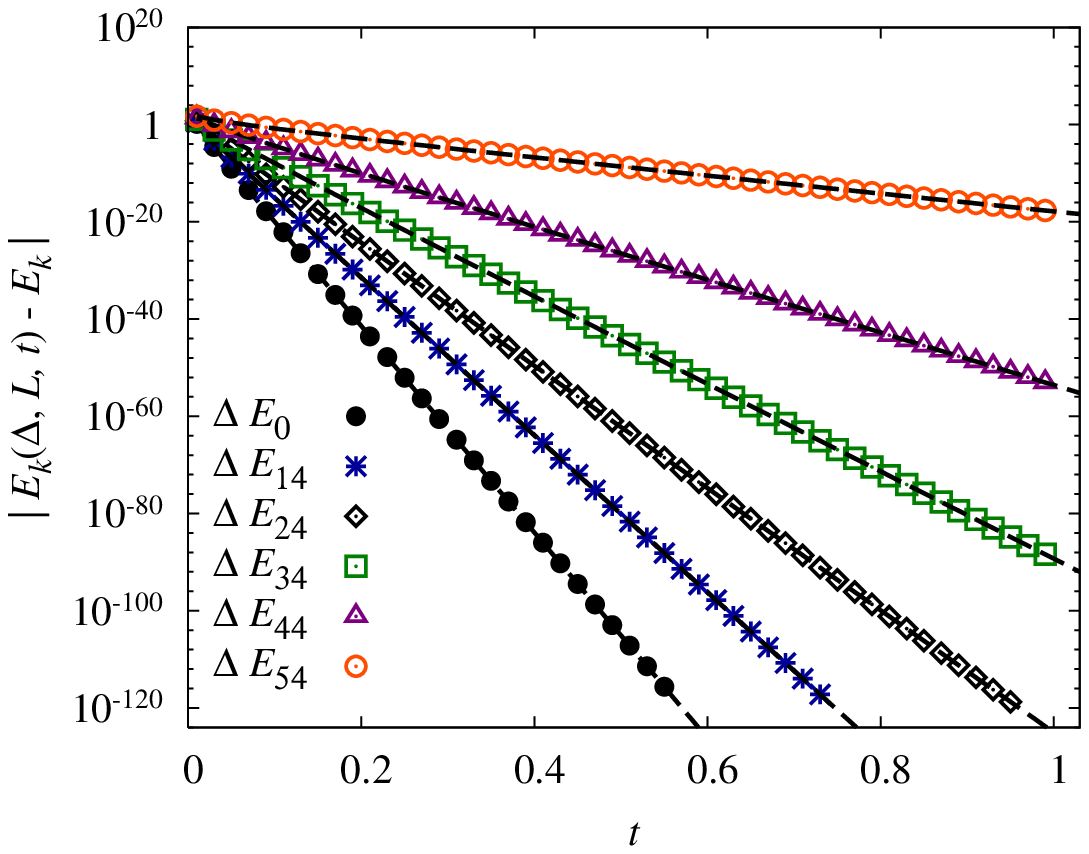}
\caption{(Color online) (a) Plot of $|E_0(\Delta,L,t)-E_0|$ for a free particle in a box as a function of $\Delta$ for different values of time of evolution $t$ and $L=6$. For comparison, we also plot the corresponding deviations of numerical results (designated by $H$) obtained using direct diagonalization of the space-discretized Hamiltonian, defined by Eq.~(\ref{eq:tb}). The inset gives a closer view of direct Hamiltonian diagonalization errors, since they have much weaker dependence on the spacing $\Delta$. (b) This plot shows how the deviations $|E_k(\Delta,L,t)-E_k|$ depend on $t$ for several energy levels $k$. The parameters used are $L=6$, $\Delta=0.2$. In both plots the dashed lines represent discretization error estimates given in Eq.~(\ref{eq:disserr}).}
\label{fig:freedis}
\end{figure}

This universal and non-perturbatively small deviation from the continuum indicates that one should center numerical calculation schemes on transition amplitudes rather than the Hamiltonian. By diagonalizing the transition amplitude for any time of propagation $t$ we obtain the energy eigenvalues and eigenfunctions
\begin{equation}
\int \mathrm{d}y\, A(x,y;t)\psi_k(y)=e^{-tE_k}\psi_k(x)\ .
\end{equation}
To solve this numerically we first discretize space with discretization step $\Delta$, and second we introduce a spatial cut-off $L$ such that $|x|<L$. Amplitudes are now
$2N\times 2N$ matrices whose diagonalization leads to
$2N$ eigenstates $\psi_k$ and eigenvalues $e^{-tE_k(\Delta,L,t)}$.

As we have seen, discretization introduces a non-perturbatively small error in transition amplitudes proportional to $\exp\left(-2\pi^2t/\Delta^2\right)$. We should therefore expect the discretization error for energy eigenvalues to be
\begin{equation}
E_k(\Delta,L,t)-E_k\sim -\frac{1}{t}\,\exp\left(-\frac{2\pi^2}{\Delta^2}t\right)\ .
\label{universal}
\end{equation}
We have numerically investigated this for a diverse set of models and have shown the above relation to hold in all cases. It is also illustrative to verify this for analytically tractable models. Using the known analytical expressions for transition amplitude and energy eigenstates for a free particle in a box \cite{kleinertbook,pib}, as well as the Poisson summation formula in Eq.~(\ref{poisson}), we find that the energy eigenstates of the space discretized model satisfy
\begin{equation}
E_k(\Delta,L,t)-E_k=-\frac{2}{t}\,e^{-\frac{2\pi^2}{\Delta^2}t}\cosh\left(\frac{\pi^2 (k+1)t}{L\Delta}\right)\ ,
\label{eq:disserr}
\end{equation}
where $E_k=\frac{\pi^2(k+1)^2}{8L^2}$ and $k=0,1,2,\ldots$
As expected, the universal term gives the dominant $\Delta$ dependance. One obtains similar analytical results for the case of the harmonic oscillator.

The non-perturbatively small effect of spatial discretization is the reason why the new method highly outperforms direct diagonalization of the Hamiltonian and leads to much smaller errors for the same size of discretization step $\Delta$. In addition to this the free parameter associated with the method, the time of evolution $t$, can be used to further minimize errors. As illustrated in Fig.~\ref{fig:freeeigenspectrum}, while keeping $\Delta$ fixed, we can adjust time $t$ to obtain much smaller errors and practically reproduce the exact spectrum of the theory. This is also evident in Figs.~\ref{fig:freedis}a and \ref{fig:freedis}b, where we see that by adjusting $t$, errors can be reduced by orders of magnitude for fixed value of discretization step $\Delta$.

\begin{figure}[!b]
\centering
\includegraphics[width=8.5cm]{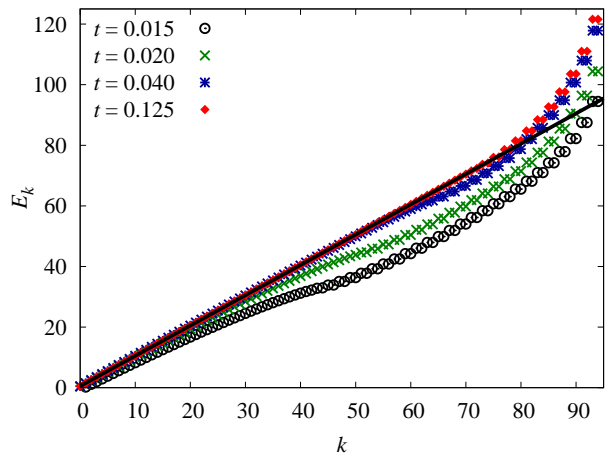}
\caption{(Color online) Harmonic oscillator dispersion relation. The solid line gives the exact linear dispersion $E_k=k+1/2$. The points correspond to numerically calculated energy eigenvalues $E_k$ as function of level $k$. We show the results of the diagonalization of transition amplitudes for several values of $t$. In this plot $L=12$, $\Delta=0.25$, the frequency is $\omega=1$, mass $M=1$.}
\label{oscillatorspectrum}
\end{figure}
\begin{figure}[!t]
\centering
\includegraphics[width=8.5cm]{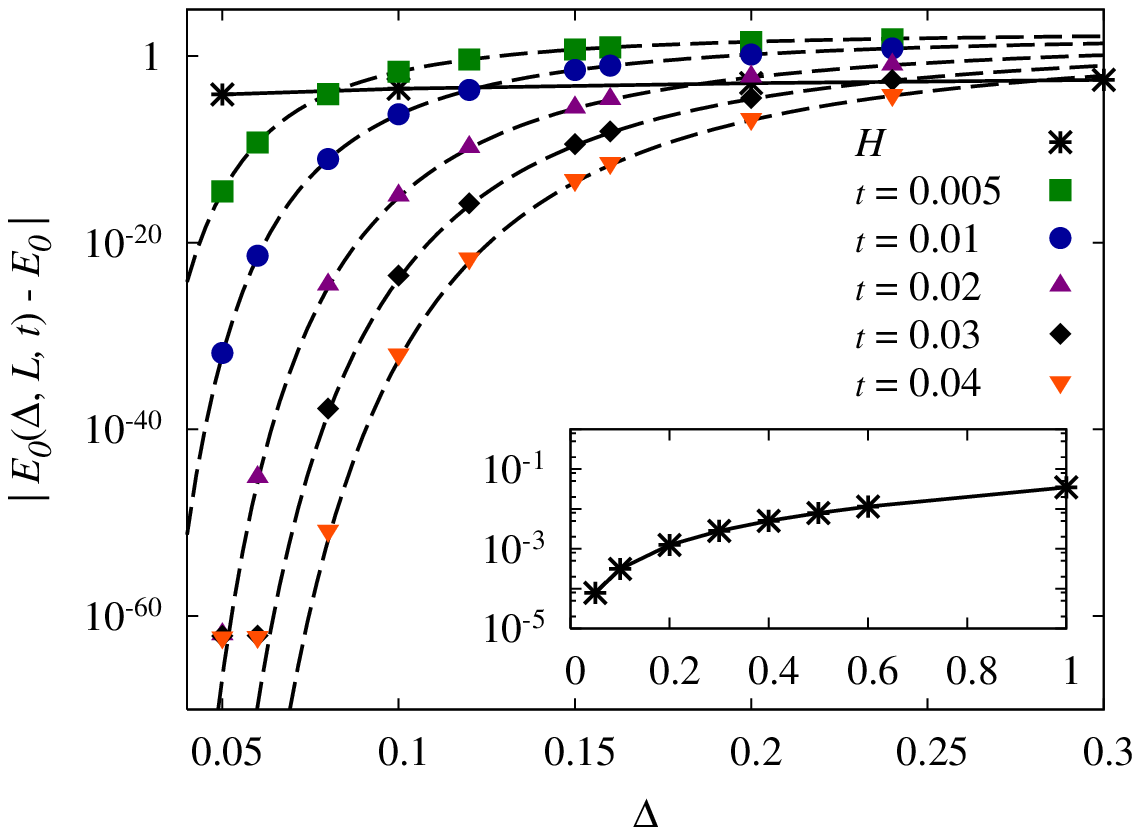}
\includegraphics[width=8.5cm]{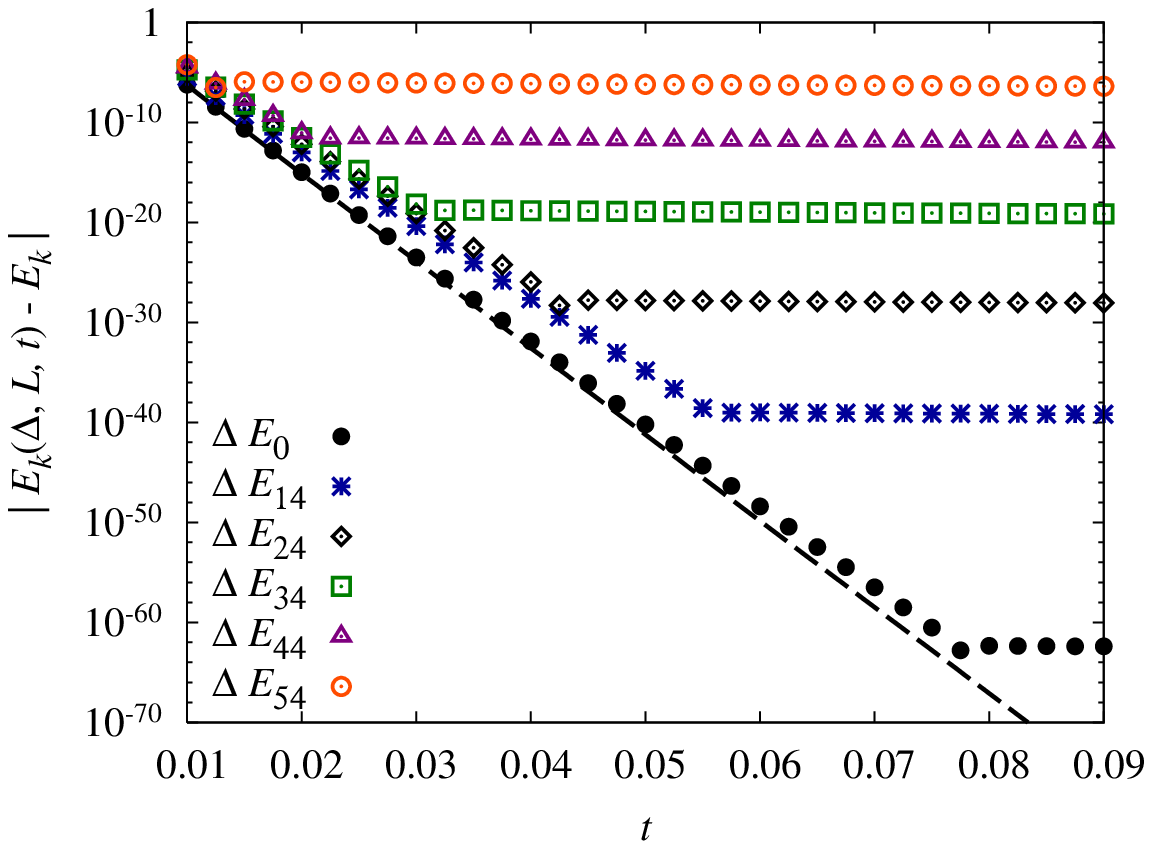}
\caption{(Color online) (a) Plot of $|E_0(\Delta,L,t)-E_0|$ for  a harmonic oscillator as a function of discretization step $\Delta$ for different values of time of evolution $t$ with $L=12$, $\omega=1$, and $M=1$. For a comparison, we also plot the corresponding results (designated by $H$) obtained using direct diagonalization of the space-discretized harmonic oscillator Hamiltonian. The inset gives a closer view of direct Hamiltonian results, since they have much weaker dependence on the discretization step $\Delta$. (b) Plot of the deviations $|E_k(\Delta,L,t)-E_k|$ given as a function of time $t$ for several levels $k$. The parameters used are $L=12$, $\Delta=0.1$, $\omega=1$, $M=1$. The observed saturation of errors for large $t$ comes about when the discretization error becomes smaller than the error due to space cutoff $L$. In both plots the dashed lines correspond to the discretization error estimate for $E_0$ given in Eq.~(\ref{eq:disserr}).}
\label{oscillatordeviations}
\end{figure}

We next consider a harmonic oscillator. Fig~\ref{oscillatorspectrum} shows how the presented method may be used to obtain energy eigenvalues to high levels. The numerical calculations agree well with the well known linear dispersion of the harmonic oscillator. Figs.~\ref{oscillatordeviations}a and b display respectively the $\Delta$ and $t$ dependance of the deviations $|E_k(\Delta,L,t)-E_k|$, showing agreement with the analytically derived estimate of the discretization error given in Eq.~(\ref{universal}). In order to achieve such a high accuracy of numerical results as presented on all graphs, we have used the Mathematica software package \cite{mathematica}.

Fig.~\ref{oscillatordeviations}b shows how the deviations $|E_k(\Delta,L,t)-E_k|$ depend on $t$ for several levels $k$. The plot corresponds to the harmonic oscillator but is typical of a general theory. The saturation of errors for large $t$ comes about when the discretization error, given by the universal estimate in Eq.~(\ref{universal}), becomes smaller than the error due to space cutoff $L$. Analytical estimates for cutoff error are given at the end of this section. At this point we just mention that the finite size effects can already be seen in Fig.~\ref{oscillatorspectrum} where for high values of level number $k$ numerical results start to move away from the linear dispersion characteristic of a harmonic oscillator to the parabolic dispersion characteristic of a box potential.

So far we have considered only integrable models, i.e. models for which we know the exact transition amplitudes. As a result we have thus far encountered and analyzed only two sources of errors: those associated with discretization step $\Delta$ and cutoff $L$. The vast majority of models are not integrable. The outlined method is still applicable if one uses some approximation for calculating transition amplitudes. In a previous series of papers we have used the method of effective actions to calculate short time expansions of transition amplitudes of a general theory to high order $p$.
For the case of a general many particle theory in arbitrary number of dimensions we have obtained closed form expressions for expansions up to $p=10$. The analytical procedure is substantially simplified for certain potentials. In particular, for polynomial interactions we have obtained expressions to level $p=144$. This high level of precision makes these short time expansion formulas ideal for use in the method outlined in this paper. Still, the diagonalization of approximate transition amplitudes introduces a third source of error proportional to $t^p$.

The overall error is minimized when all three sources of error are approximately equal. The key point is that we have simple analytical estimates for all three errors. The universal behavior of the discretization error substantially simplifies the process by which one chooses the values of parameters $\Delta$, $L$, and $t$ that minimize the overall error. The details of this are presented in the followup paper.

\begin{figure}[!t]
\centering
\includegraphics[width=8.5cm]{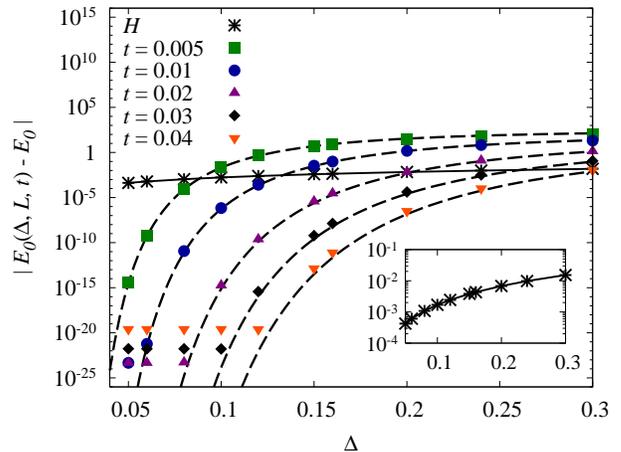}
\caption{(Color online) Plot of $|E_0(\Delta,L,t)-E_0|$ for an anharmonic oscillator with potential $V=\frac{1}{2}M\omega^2 x^2+\frac{g}{24}x^4$ given as a function of $\Delta$ for different values of time of evolution $t$. The parameters used are $L=6$, harmonic frequency $\omega=1$, mass $M=1$, and anharmonicity $g=48$. Transition amplitude matrix elements were calculated using $p=18$ effective actions \cite{b}. The high precision value for the exact ground energy that we compare to was calculated in Ref.~\cite{c}. Dashed lines correspond to the discretization error in Eq.~(\ref{eq:disserr}). For comparison, we also plot the corresponding deviations of numerical results (designated by $H$) obtained using direct diagonalization of the corresponding space-discretized Hamiltonian. The inset gives a closer view of direct Hamiltonian diagonalization errors, since they have much weaker dependence on the discretization step $\Delta$.}
\label{anharmonic}
\end{figure}

Fig.~\ref{anharmonic} displays $|E_0(\Delta,L,t)-E_0|$ as a function of discretization step $\Delta$ for the case of an anharmonic oscillator with potential $V=\frac{1}{2}M\omega^2 x^2+\frac{g}{24}x^4$. The parameters used in the plot are $L=6$, $\omega=1$, $M=1$, and anharmonicity $g=48$. The transition amplitude matrix elements were calculated using $p=18$ effective actions \cite{b}. The high precision value for the exact ground energy that we compare to was calculated in Ref.~\cite{c}. As we can see, even though we are dealing with a relatively strong anharmonicity, the numerically calculated values stay right on the dashed lines corresponding to the universal discretization error just as in the case of the previously considered integrable models. This is in complete agreement with our analytical derivation of the discretization error. In the followup paper we numerically investigate a  variety of different interacting models and in all cases document the validity of this formula.

As can be seen from Fig.~\ref{anharmonic}, the numerical results clearly demonstrate that the $\Delta$-dependence of errors within our calculation scheme highly outperforms the polynomial dependence in $\Delta^2$ obtained by the direct diagonalization of the Hamiltonian. This is true even for short times of propagation $t$. Although interaction terms in the potential affect the numerical values of errors, diagonalization of the transition amplitudes still substantially outperforms diagonalization of the Hamiltonian, and is the preferred method. This success is a consequence of the non-perturbative behavior of the spatial disretization error within this calculation scheme. This leads us to the key conclusion that discretization parameters can be always optimized so that presented approach of solving eigenvalue problem of space-discretized transition amplitudes highly outperforms direct diagonalization of the space-discretized Hamiltonian. The continuum limit $\Delta\to 0$ is far more easily approached in the first case and the corresponding discretization errors are substantially smaller for  the same discretization coarseness. From the numerical point of view, as the value of parameter $\Delta$ directly determines the size of the matrix to be diagonalized, the computational cost for the same precision is significantly reduced.

We end by looking at finite size effects, i.e. errors related to introduction of space cutoff $L$. For any theory with non-trivial potential, the cutoff $L$ is artificially introduced and it affects the obtained energy eigenvalues, as we have already discussed. To estimate the effects of the cutoff, we first note that they are closely related to the spatial extent of the potential $V$, as well as the spatial extent of eigenfunctions of the system: errors in the corresponding energy eigenvalues can be considered small only if the eigenstates $\psi_k(x)$ are well localized in the interval $|x|<L$.

The effects of space cutoffs have been previously studied for continuous-space theories \cite{ajp1, ajp2}. The shift in energy level $E_k(L)-E_k$ is found to be positive in this case, and approximately given by the formula
\begin{equation}
E_k(L)-E_k=C_k(a)\left(\int_a^L\frac{\mathrm{d}x}{|\psi_k(x)|^2}\right)^{-1}\, ,
\label{eq:DeltaE}
\end{equation}
where $a$ is an appropriately chosen value of coordinate $x$ such that it is larger than and well away from the largest zero of $\psi_k(x)$ but smaller than and well away from the space cutoff $L$. The constant $C_k(a)$ depends on the normalization of eigenfunction and the choice of parameter $a$. For example, the ground state has no zeros, and we can always choose the value $a=0$. In that case, constant $C_0(0)$ is given by
\begin{equation}
C_0(0)=\left(\int_{-L}^L \mathrm{d}x\,|\psi_0(x)|^2\right)^{-1}\, ,
\end{equation}
where we assume that the eigenfunction $\psi_0(x)$ is normalized as usual, $\int_{-\infty}^\infty \mathrm{d}x\,|\psi_0(x)|^2=1$.

\begin{figure}[!t]
\centering
\includegraphics[width=8.5cm]{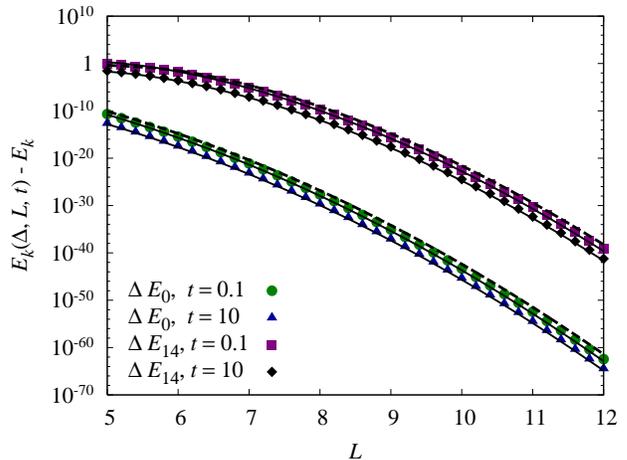}
\caption{(Color online) Deviations $E_k(\Delta,L,t)-E_k$ for a harmonic oscillator as a function of space cutoff $L$ for different values of time of evolution $t$. The parameters used are $\Delta=0.1$, $\omega=1$, $M=1$. Solid thin lines give the dominant behavior of Eq.~(\ref{eq:varerr}). The dashed thick lines correspond to the error estimate in Eq.~(\ref{eq:DeltaE}).}
\label{fig:lhoLpodaciv1}
\end{figure}

In practical applications, when we use diagonalization of the discretized transition amplitudes, the errors in energy level will necessarily also depend on the parameter $t$ and other discretization parameters. Here we give a simple estimate of ground energy errors that follows from the spectral decomposition of diagonal amplitudes. For large $t$ we have $A(x,x;t)\approx |\psi_0(x)|^2 e^{-E_0 t}$. Integrating this we find an approximate result for the ground energy of a system with cutoff $L$
\begin{equation}
 E_0(L,t)\approx -\frac{1}{t} \ln \int_{-L}^{L} \mathrm{d}x\,A(x,x;t)\, ,
\label{eq:e0ap}
\end{equation}
In the $L\to\infty$ limit we recover the exact ground energy, so that a simple estimate of finite size effects on $E_0$ is given by
\begin{equation}
E_0 (L, t)- E_0\approx \frac{1}{t} \int_{|x|>L} \mathrm{d}x\,|\psi_0(x)|^2\, .
\label{eq:varerr}
\end{equation}
Although the above equation is just a rough estimate of the errors introduced by a space cutoff $L$, Fig.~\ref{fig:lhoLpodaciv1} shows that it is in good agreement with numerical results for the harmonic oscillator. In order to clearly demonstrate $L$-dependence of errors in this graph, we have used small value of the discretization step $\Delta$, such that discretization errors can be neglected. The dashed line in the figure represents error estimates given by Eq.~(\ref{eq:DeltaE}).

Using the data from Fig.~\ref{fig:lhoLpodaciv1} we can now fully explain the saturation of errors observed in Fig.~\ref{oscillatordeviations}b. The value of the cutoff $L$ used to obtain this data was $L=12$. As can be seen from Fig.~\ref{fig:lhoLpodaciv1}, this value of the cutoff parameter yields an error of the order $10^{-65}$ for the ground-state energy for $t\sim 0.1$, and of the order $10^{-40}$ for energy eigenlevel $E_{14}$. These values exactly correspond to the saturated errors in Fig.~\ref{oscillatordeviations}b.

Although in the general case the eigenstates that come into Eqs.~(\ref{eq:DeltaE}) and (\ref{eq:varerr}) are not known, we can still use them in conjunction with other approximation techniques to estimate finite size effects. We also see that, due to the larger spatial extent of higher energy eigenstates, the cutoff-related errors are minimal for the ground energy. Note however that one is not really interested in the precise calculation of finite size errors, but only needs to estimate the minimal size of the cutoff $L$ for which finite size effects are negligible. For that purpose one can use either of the above approximate formulas.

\section{Conclusions}
\label{sec:conclusions}

The current paper is the first in a series of two publications dealing with the properties of quantum systems calculated from the the diagonalization of transition amplitudes. In this paper we have focused on analyzing the errors associated with real-space discretization and finite size effects. In particular, we have shown that within this calculation scheme spatial discretization leads to a universal and non-perturbatively small discretization error. This highly outperforms the usual polynomial behavior of errors in approaches based on the diagonalization of space-discretized Hamiltonians. In addition to providing a full understanding of the numerical method based on diagonalization of the evolution operator, we have also derived analytical estimates for all the errors involved within this approach. In practical applications, the derived analytical results make it possible to optimize parameters of the method so as to minimize errors in calculated energy eigenvalues and eigenstates in the case of a general theory.

The second paper in the series \cite{pqseeo2} builds on these results, extending them through systematic improvement of short-time propagation using the effective action approach \cite{b}. This effectively solves the problem of the accurate calculation of evolution operator matrix elements and significantly reduces errors related to the time of evolution parameter.

\section*{Acknowledgments}
The authors would like to acknowledge helpful discussions with Antun Bala\v z. This work was supported in part by the Ministry of Science and Technological Development of the Republic of Serbia, under project No. OI141035 and bilateral project PI-BEC funded jointly with the German Academic Exchange Service (DAAD), and the European Commission under EU Centre of Excellence grant CX-CMCS. Numerical simulations were run on the AEGIS e-Infrastructure, supported in part by FP7 projects EGEE-III and SEE-GRID-SCI.


\begin {thebibliography}{00}

\bibitem{pqseeo2}
I. Vidanovi\' c, A. Bogojevi\' c, A. Bala\v z, A. Beli\' c, Phys. Rev. E {\bf 80} (2009) 066706; arXiv:0911.5154

\bibitem{sethia}
A. Sethia, S. Sanyal, Y. Singh, J. Chem. Phys. {\bf 93} (1990) 7268.

\bibitem{sethiacpl1}
A. Sethia, S. Sanyal, F. Hirata, Chem. Phys. Lett. {\bf 315} (1999) 299.

\bibitem{sethiajcp}
A. Sethia, S. Sanyal, F. Hirata, J. Chem. Phys. {\bf 114} (2001) 5097.

\bibitem{sethiacpl2}
S. Sanyal, A. Sethia, Chem. Phys. Lett. {\bf 404} (2005) 192.

\bibitem{b}
A. Bala\v z, A. Bogojevi\' c, I. Vidanovi\' c, A. Pelster, Phys. Rev. E {\bf 79} (2009) 036701.

\bibitem{prl-speedup}
A. Bogojevi\' c, A. Bala\v z, A. Beli\' c, Phys. Rev. Lett. {\bf 94} (2005) 180403.

\bibitem{prb-speedup}
A. Bogojevi\' c, A. Bala\v z, A. Beli\' c, Phys. Rev. B {\bf 72} (2005) 064302.

\bibitem{pla-euler}
A. Bogojevi\' c, A. Bala\v z, A. Beli\' c, Phys. Lett. A {\bf 344} (2005) 84.

\bibitem{pla-manybody}
A. Bogojevi\' c, I. Vidanovi\' c, A. Bala\v z, A. Beli\' c, Phys. Lett. A {\bf 372} (2008) 3341.

\bibitem{k1}
S.~A. Chin, S. Janecek, E. Krotscheck, Comp. Phys. Comm. (2009), In press, doi:10.1016/j.cpc.2009.04.003

\bibitem{k2}
S. Janecek, E. Krotscheck, Comp. Phys. Comm. {\bf 178} (2008) 835.

\bibitem{k3}
M. Aichinger, S.~A. Chin, E. Krotscheck, Comp. Phys. Comm. {\bf 171} (2005) 197.

\bibitem{spacedisrev}
T.~L. Beck, Rev. Mod. Phys. {\bf 72} (2000) 1041.

\bibitem{ajp1}
G. Barton, A.~J. Bray, A.~J. McKane, Am. J. Phys. {\bf 58} (1990) 751.

\bibitem{ajp2}
D.~H. Berman, Am. J. Phys. {\bf 59} (1991) 937.

\bibitem{d}
M.~A. Martin-Delgado, G. Sierra, R.~M. Noack, J. Phys. A {\bf 32} (1999) 6079.

\bibitem{spacedis1}
J.~R. Chelikowsky, N. Troullier, K. Wu, Y. Saad, Phys. Rev. B {\bf 50} (1994) 11355.

\bibitem{spacedis2}
P. Maragakis, J.~M. Soler, E. Kaxiras, Phys. Rev. B {\bf 64} (2001) 193101.

\bibitem{skylaris}
C.-K. Skylaris, O. Di\' eguez, P. D. Haynes, M.~C. Payne, Phys. Rev. B {\bf 66} (2002) 073103.

\bibitem{kleinertbook}
H. Kleinert, Path Integrals in Quantum Mechanics, Statistics, Polymer Physics, and Financial Markets, 4th edition, World Scientific, Singapore, 2006.

\bibitem{pib}
W. Janke, H. Kleinert, Lett. Nuovo Cimento {\bf 25} (1979) 297.

\bibitem{mathematica}
Mathematica symbolic calculation software package, {\tt http://www.wolfram.com/}

\bibitem{c}
W. Janke, H. Kleinert, Phys. Rev. Lett. {\bf 75} (1995) 2787.


\end{thebibliography}
\end{document}